\documentclass{article}

\newcommand{\bc}{\begin{center}}
\newcommand{\ec}{\end{center}}
\newcommand{\bd}{\begin{displaymath}}
\newcommand{\ed}{\end{displaymath}}
\newcommand{\be}{\begin{equation}}
\newcommand{\ee}{\end{equation}}
\newcommand{\ba}{\begin{array}}
\newcommand{\ea}{\end{array}}
\newcommand{\bea}{\begin{eqnarray}}
\newcommand{\eea}{\end{eqnarray}}
\newcommand{\bt}{\begin{tabular}}
\newcommand{\et}{\end{tabular}}

\newcommand{\bp}{\begin{picture}}
\newcommand{\ep}{\end{picture}}
\newcommand{\bfi}{\begin{figure}}
\newcommand{\efi}{\end{figure}}

\begin{document}


\title{\huge \bf { Remarkable Relation from Minimal Imaginary Action Model
\footnote{This article is essentially to be considered a proceeding 
contribution to the Spaatind Nordic Particle Physics Meeting 2010, in which
I delivered a talk ``Playing Cards with ``God'' ''.  }  }}

\author{ 
H.B.~Nielsen ${}^{1}$ \footnote{\large\, hbech@nbi.dk} \\[5mm]
\itshape{${}^{1}$ The Niels Bohr Institute, Copenhagen, Denmark}}

\date{}

\maketitle

\begin{abstract}
If it is to be true that the history of the universe should minimize the 
``imaginary part of action'' \cite{own} \cite{degenerate}  it would 
be ``easiest'' to have 
this ``imaginary 
part of the action'' to be rather independent of whether the neutrons are 
converted to electrons and protons (and neutrinos), and therefore of whether 
the 
contribution to the imaginary part of the Lagrangian were conserved 
under this convertion. Under the further assumption - which is reasonable - 
that the by far dominant term in the imaginary part of the Lagrangian density 
is the one corresponding to the Higgs mass square term in the Standard 
Model Lagrangian density $m_H^2 |\phi(x)|^2$ we derive a relation 
\begin{equation}
\sqrt{m_d^2-m_u^2} = \sqrt{m_{constituent}*m_e}/\sqrt{``ln''}
\end{equation}    
where $``ln''$ is defined  by $<\gamma^{-1} > = ``ln''/<\gamma>$ and is of the 
order of  1 to 4, where $\gamma$ stands for relativity $\gamma$ of a valence 
quark inside a nucleon. 
 This relation is very well satisfied with the 
phenomenologically estimated current algebra quark masses $m_u$, $m_d$, 
and the constituent mass $m_{constituent}$ for the light quarks, taken say 
to be one third of the nucleon mass.

Our model has been criticised on the ground that it should have prevented 
cosmic rays 
with energies capable of producing Higgs particles, when hitting say the 
atmosphere. Indeed there is, however, a well known ``knee'' in the 
density
curve as function of energy at an order 
of magnitude for the cosmic ray particle energy close to the effective 
threshold for Higgs production.

A parameter giving the order of magnitude of number of Higgsses to be 
produced in order to get a significant effect is estimated to be 
about $3*10^5$.        
\end{abstract}

\newpage
\thispagestyle{empty}
\section{Introduction}
Recently Ninomiya and I \cite{own}\cite{degenerate} have put forward after 
some thinking on 
time machines \cite{ownctl} a model 
in which the action - of the Standard Model say - gets provided with an 
imaginary part. 
This having a model for the {\em initial conditions} is 
also
a subject already touched upon by me and Bennett much earlier 
\cite{vacuumbomb}\cite{nonlocal}.  It were suggested that the term in the 
Lagrangian density 
corresponding Higgs mass square $m_H^2 |\phi(x)|^2$ should dominate the 
imaginary part and that leads to the main effect of the model being to 
predict that accelerators producing large amounts of Higgs particles 
should be likely to have bad luck in the sense of not coming to fully
work.

It is a major point of the present article to point out that with the actually 
natural assumption of the Higgs mass square term dominating  the number of 
parameters in the imaginary part of the action is formally reduced to the 
imaginary part $m_H^2|_I = Im(m_h^2)$ of the mass square term coefficient.
In reality the exact variation of the integrated square of the Higgs field 
might be not completely easy to evaluate and a few parameters parametrizing 
the difficulties in performing calculations which in principle can be done 
might have 
to be included into the model. 

We have in earlier works already argued, that for a situation with 
only massless and conserved non-relativistically moving particles 
the effects of the situation on the imaginary action $S_I(history)$
become trivial and thus there will be in this approximation no effect 
of the requirement predicted by our model for governing 
the initial conditions of the universe so as to arrange that 
what really happens minimizes this imaginary part of the action 
$S_I(history)$. 

We should thus only make unusual predictions when either 
\begin{itemize}
\item{a)} Some particles are not either massless or non-relativistic,
(In that case namely a more complicated form for the eigentime for such 
particles would appear, than for massless or non-relativistic particles 
with the eigentimes being respectively zero and equal to the usual coordinate 
time; more complicated eigentime would mean also more complicated action,
say the imaginary art, because the action goes as the eigentime.)   
or 
\item{b)} Some particles are not conserved, but say converted into each 
other or simply appear or disappear.(Such decay or appearance would of course 
make the to the action connected eigentime depend on when the appearance 
and decay occurs so that the history would influence the (imaginary) action 
under such conditions) 

\end{itemize}     

The point of the trivial contributions to the ``imaginary part of the 
action'' when the two mentioned deviations from 
daily life physics are avoided
is easily understood by noting that he action - it being the real of the 
imaginary part - of a propagating particle, say from one interaction to the 
next, is a constant times the (relativistic)eigentime:

We namely then immediately see that at least between the interactions 
- and interactions will in daily life physics take up a relatively short time
compared to the free times - the massless particles contribute zero because 
their eigentimes are zero, and the non-relativistically conserved particles 
contribute just proportional to the frame time in the frame in which 
they are non-relativistic, provided of course that they are conserved 
so that they are indeed present at all times.

In the so to speak daily life approximation we only meet protons, 
neutrons, electrons and massless particles meaning mainly the photon.
Among these particles only the nucleons attain velocities approaching 
relativistic speeds inside the nuclei, which themselves are only seldomly 
converted into each other. As long as the nuclear reactions do not occur 
and the motions of the particles remain non-relativistic the imaginary 
action contributions remain trivial to good approximation and severe 
governing of the universe effect such as some apparatuses having ``bad 
or good luck'' should not occur.

Astronomically there does, however, occur some transformations of 
electrons plus protons into neutrons and netrinos. Since the neutrinos 
are to first approximation massless they should not contribute to the 
imaginary part of the Lagrangian for that reason. But a priori, unless 
the contributions to the imaginary part of the Lagrangian of a neutron 
were almost miraculously tuned in to be the same as that from the 
electron and the proton together there would be a significant change 
in the imaginary part of the lagrangian $L_i$ each time the reaction 
\begin{equation}
e+p \leftrightarrow  n + \nu_e \label{wp}
\end{equation}  
runs one way or the other. 

It is the main point of the present article to argue for that it could 
be likely that the minimization of $S_I$ would lead to a tune in of parameters 
that would organize this 
transition  $e+p \leftrightarrow  n + \nu_e$ to {\em not} change the $L_I$. 
The idea 
is in the direction that other features of the initial conditions can be 
adjusted more freely without having to take care of their influence 
on the amount of transitions of this type of 
``neutron decay'' taking place or at what time they take place.  

For the argument for this sort of finetuning being likely to occur we refer 
to section \ref{need}.

\section{Review of our Imaginary part of Action Model}

One may formulate the model by M. Ninomiya and myself \cite{own} by 
saying that we use a certain way of interpreting the Feynman-Dirac-Wentzel
path way integral (a  bit different from the usual use) we assume that the 
action $S[path]$ has also {\em an imaginary part}. This means of course 
that we write the action $S$ to put into the Feynman path way integral
\be
\int exp(\frac{i}{\hbar}*S[path]){\cal D}path\label{amplitude}
\ee 
as
\be
S[path] = S_R[path] +i S_I[path]
\ee
where of course $S_I[path]$ is the imaginary part, which is zero in the 
usual theory, but assumed to be a functional of quite similar, but not 
identical, form to that of the real part $S_R[path]$. 

It is immediately obvous that, while in the usual theory in which there is 
no imaginary part of the action, i.e. $S_I[path]=0$, the integrand of the 
path way integral $  exp(\frac{i}{\hbar}*S[path]) $ is a complex number 
with unit numerical value, the inclusion of our imaginary part makes 
the integrand able to take a pathdependent numerical value. In fact it is 
seen that assuming the imaginary part to be a priori of a similar order of 
magnitude as the real part of the action and remembering that compared to 
daily life, when classical approximation is often good enough, the Planck 
constant $\hbar$ is in principle small the suppression of the integrand 
for positive $S_I[path]$ and the enhancement for negative $S_I[path]$ are
enormous. Without further details this clearly points towards the idea 
that the paths which do not have their $S_I[path]$ being minimal inside 
the range of paths not brought to be made irrelevant for other reasons 
must come to dominate the path way integral.

Now it is wellknown that one can in the usual path way integral formulation 
argue for the classical equations of motion to appear as a saddle point 
approximation in the evalutaion of the Wentzel-Feynman-Dirac path way integral.
Basically the idea in the derivation from the path way integral of the 
classical equations of motion comes from that unless we sum up the 
different path contributions in the neighborhood of a path for which 
the variation of the action is zero
\be
\delta S[path] =0
\ee
the phase of the integrand will vary over the neighboring region 
of that path so fast, that essentially the contributions cancel out, and we 
get very little resulting contribution from the neighborhood of such a path.

Very crudely we might take this phase variation argument to mean that only 
the neighborhood of paths for which 
\be
\delta S_R[path] =0
 \ee
will survive (i.e stay making a significant contribution to the path way 
integral) and that then the path{\em  which among these classical solutions}
have the minimal value for $S_I[path]$ will dominate. This approximation 
strictly speaking 
does not work when the $S_I$ gets big, but we believe to have arguments 
that it will work approximately in practice. 

\subsection{We include the future}
A most important point in our way of applying the path way integral 
compared to what is usually done is that we {\em include the contributions 
from the future} into the integral form of the action
\be
S[path] = \int_{\infty}^{\infty} L(path(t), \frac{d path(t)}{d t}) dt. 
\ee 

This inclusion of the future has the very important consequence that 
the question as to which path inside, say the class of those ones  obeying the 
classical equations of motion,  gets the minimal value of the imaginary 
part of the action $S_I[path]$ and thus should be the most important path
depends not only on what goes on along this path in the past, but 
also {\em in the future}. That has the consequaence that it will look 
in our model as if the initial conditions have been {\em selected with 
some purposes of arranging in the future some happenings that could make 
especially numerically big but negative contribtuions to $S_I[path]$}.
Such prearrangements to obtain especially negative contributions to $S_I$
even from the future will be concieved of as {\em miraculous coincidences 
seemingly made with a purpose.} Provided our speculation that especially 
the production of Higgs bosons in large numbers should cause exceptionally 
large positive constributions to the imaginary part of the action $S_I$
we expect the typical miraculous coincidences of this type to be the 
coincidenses of small bad lucks leading to that a great accelerator meant 
to produce a lot of Higgs bosons after all gets in trouble and does not at 
the end come to work. We have since long proposed that the case of the 
stopping 
in 1993 of the budgets for the SSC (= Superconducting Supercollider in Texas)
by the U.S. Congress were a case of such a prearranged set of coincidences 
``made up'' to stop the potential great amount of Higgs bosons being 
produced.   

\subsection{The classical approximation law of initial conditions}
As we have just stated the bunch of paths that will dominate the 
Feynman-Wentzel-Dirac path integral in our model with its complex 
action is expected to be
\begin{itemize}
\item{1)} a bunch around a classical solution (since otherwise 
the phase from the integrand corresponding to the different paths 
in the bunch would vary so as to make the contribution of the bunch wash
out)

\item{2)} the bunch of this type having the minimal - in the sense of 
being most negative - imaginary part of the action $S_I(history)$.  
\end{itemize}

Supposing that the point 1) can be apprimated by just including the 
real part of the action and taking the usual classical equation of motion 
written in the usual extremizing the action way,
\begin{equation}
\delta S_R =0,
\end{equation}
the point 2) takes formally the form of a minimization among all the 
classical solutions
\begin{eqnarray}
 & LAW \;\; OF \;\; INITIAL \; CONDITIONS :&  \\
 & S_I(history) \;\; MINIMAL.&
\end{eqnarray}
This is much like a law of ``the will of ``God'' ''.

\subsection{Suppression of the effect of the imaginary part $S_I$}
Both for the performance of the argument for the main effect   
indeed being of the form that the classical solution selected to be 
realized is the one with the minimal imaginary part of the action $S_I$
and for obtaining agreement with the phenomenological fact that after all
we see extremely few - if at all any - prearranged events we need for our model
to be viable that in some way effectively the imaginary is small. 
By prearranged event we here meant 
that we should see something  requiring an explanation involving special 
finetuning of initial conditions so as to arrange some special thing to 
happen later. Even if we would count the failure of the SSC machine as such 
a case and include some miracles from the bible etc. such events - which 
might be called cases of backward causation - are extremely rare.

So we have in our model the problem that we either have to assume that 
even if there is an imaginary part of the action then it is very small,
or we have to presnt a mechanism, that even if there is a priori say 
an imaginary part of the action being of the same order as the real one,
then in pracsis we shall see effects only as if it were much smaller.

\subsubsection{An important argument for suppressing the effect of the 
imaginary part of the action}

The argument for suppressing the effect to which we have most hope 
is of the type that each periode of time, each era, cannot have very much 
say about the initial conditions, because there is only one set of initial 
conditions to determine what happens in all the many eras through the 
history of the Universe:

One may see each era in the development of the Universe as having different
developments, all being though described according to the equations of 
motion being integrated up from the same initial conditions. It is the same 
equation of motion solution that must describe what goes on in all eras 
from the earliest times to the latest times. Since the solutions are so to 
speak in one to one correspondance with the state (in phase space) at one 
moment of time there are of course not freedom enough to adjust for the 
optimal (meaning $S_I$ being minimal) happening for all the many eras 
with the same solution. Rather the minimization of the imaginary part of the 
action $S_I(history)$, which is an intergal over all times, must be determined
as a compromise between ``wishes'' from the many different eras. But that will
then in each seperate era make events happen that do not precisely make the 
$S_I$-contribution from just that era become minimal. Rather it would look from
the point of view of one single era as if what happens is mainly determined 
from the influence on the choice of solution from quite different eras.
Really to make our model match with phenomenological facts we must hope 
that for some reason in our model the era of the big bang time contributes
especially strongly to $S_I$ so that what we observe will in first 
approximation look like being determined by organization of an arrangement 
of what went on in the Big Bang time(s). Thereby it will namely come to 
look as if the past is fixed so as to fullfill some - especially negative 
$L_I$ - story during that era, and then we mainly just ``see'' the 
essentially unavoidable development of this in a big bang era fixed 
solution to the equations of motion.           
   
\section{The need for $L_I$ being insensitive to  Neutron decay or its 
inverse}\label{need}

\subsection{Selection of initial state may in many cases lead to selection 
of coupling constants in pracsis!}

A priori our model predicts that the minimization of the imaginary part 
$S_I(history)$ occurs by selecting the $history$ by selecting the initial 
conditions mainly. Contrary to that we at first should think of the 
coupling constants and parameters such as masses as being fixed by nature 
in a different way, a priori. However, there are ways in some models such 
as baby-universe theory that initial condition given quatities could 
achieve to influence the world effectively as if the parameters as 
coupling constants etc. were dependent on the initial conditions and thus in
our model also should be adjusted - at least to some extend - to minimize 
$S_I$, to some extend as if this $S_I$ {\em also} depended on some 
parameters parametrizing the coupling canstants and mass-parameters.
  
In arguing for how the parameters, such as coupling constants masses etc., 
might be adjusted to the degree that they are accessible to being changed in 
the 
proceedure of minimizing ``the imaginary part of the action''
$S_I[history]$ we could imagine a picture, that the vacuum has a very 
complicated structure with many fields, that could be adjusted from the start 
and then would stand so further on. Then namely the adjustment of the 
initial situation w.r.t. these ``parameters of the vacuum'' could cause 
an effective way of getting the minimization of $S_I$ also determine 
- still may be within some restrictions though - the coupling constants 
and mass parameters etc.  

The obvious example of that kind of adjustable vacuum would in pracsis be 
the ``landscape'' ideas\cite{landscape}: In superstring theory there are so 
many ways of of compactifying extra dimensions in detial that one has 
a huge number of possible effective theories in the low energy fourdimensional 
approximation. These hugely many have at least some stability so that 
the one that gets installed will stay for long time. Thus precisely which 
vacuum gets installed in the beginning will determine which vacuum and thus 
which effective couplings will be realized over later times. 

So we must imagine that in our model the vacuum to survive will be selected 
to be the one or one of the ones with the lowest imaginary part of the 
Lagrangian density $\cal \Lambda$ up to some corrections from the 
behavior of the material getting present in this vacuum. Since at least 
phenomenologically to day most of the universe is extremely empty 
one would expect that the selection of the one of the vacuua with the
lowest $\cal \Lambda$ and thereby effectively of the effective coupling 
constants for the effective fields in this vacuum would play a very important 
role in settling the initial conditions minmizing the imaginary part of the 
action $S_I$. 

But if there are indeed a very huge number of vacua to choose between it
could turn out that it would not necessarily be exactly the vacuum with the 
lowest Lagrangian density in the very vacuum situation $\cal |lambda|_{vacuum}$
which would be selected, but that a hisory with a bit lower $S_I$-contribution
from the matter and radiation etc could compete if the difference in vacuum
contribtuions is small.     


\subsection{The balance} 

Think of there being a set of parameters that are accessible to being tuned 
on to vary $S_I$ and then we know the realized choice should be one with
$S_I$ having been minimized. 
Corresponding to a set of parameters we imagine that there is then a history 
of the universe development and especially say the total number of neutrons 
existing inside or outside the nuclei at each moment of time would have some 
value depending on the history in question.
Then one could especially think of extracting for such a history  
the collected existence 
time for all the neutrons $T_{neutrons} = \sum_i t_i $ , meaning the 
sum of the time of existence $t_i$ of  all the neutrons, indexed by $i$ 
exist for some time in the world  (free or inside the nuclei in which 
there stability 
has been increased to  infinite stability  usually).
It should be emphasized that this ``existence time 
of all the neutrons   $T_{neutrons} = \sum_i t_i $ is concerned with 
mostly {\em neutrons bound into nuclei}, while free neutrons are so 
seldom in pracsis in our universe that they play practically no role in 
comparizon. We are dominantly concerned with neutrons stabilized (almost)
completlely by being bound into nuclei.

We might even think of  this collected lifetime 
of all the neutrons $T{neutrons}$ as one of the adjustable parameters, 
if we wish 
to do so. But if we think this way, then assuming a smooth behavior of 
the $S_I$ to be minimized as function of the parameters we should 
deduce that the {\em derivative} of $S_I$ w.r.t. the collected neutron 
existence  
time $T_{neutrons}$
should be {\em zero}. This is just the usual rule of the derivative being zero 
at a minimum.  Now most of the contribution to the collective neutron 
existence time comes presumably from the era which is dominated by 
the physics of ``dayly life'' as we called it above. Now when we think 
of varying the collected neutron existence time it will of course occur by 
shifting the transformation of the neutrons to proton plus electron under 
absorbsion of the neutrino or opposite earlier or later in the history
$history$. The effect of such a change in the history 
- by changing some parameters - will change the total $S_I(history)$
for two sorts of reasons:
\begin{itemize}
\item{1} There will simply be the change due to the difference in the 
$L_I$ comming from proton + electron versus that from neutron 
+ electronneutrino. 

\item{2} There will be all the other changes caused by the shift in the 
variables that were to be changed in order to arrange for the change in 
the neutron collective existence time $T_{neutrons}$ we wanted to change. 
A priori 
this change could be big, because changing the parameters will typically 
change e.g. the imaginary part of the vacuum Lagrangian density ${\cal L_I}$, 
and since there is a huge amount of vacuum this could be very big amounts 
of $S_I$-contribution. However, now we want to argue that we have the freedom 
when discussing how to produce a given little change in the collected 
neutron existence time $T_{neutrons}$ to choose which parameter to use. 
If there are many 
possible parameters to vary  - or for some reason there are great chance 
that the 
choice of parameter can lead to very small effect on $S_I$ if the parameter
is chosen appropriately -, then we could choose likely the parameter to 
play with
to cause only small influense on $S_I$ via other effects than just via the 
collected neutron existence time. 

\end{itemize}  
If we indeed take the choice among many parameters to vary when thinking of 
varying the collected neutron existence time to minimize the variation 
of $S_I$ described under point 2, then we may argue that the variation 
under point 1 will dominate. The latter is, however, {\em calculable}
in principle, since we have assumed the $L_I$ to be dominated by the 
term in the imaginary part of the Lagrangian density, which is proportional 
to the Higgs field square - we argued in fact for the imaginary coefficient 
part of the Higgs mass term to dominate-. But this then means the 
variation due to the change - as under item 1 -  in 
the collected neutron existence by itself shall be zero. Now this change in 
$S_I$
is proportional to the difference in $L_I$ for a resting (or just 
non-relativistic) proton plus a resting electron relative to 
that of a resting neutron. The neutrino is massless and we ignore its
$L_I$-contribution. Therefore the argumentation from the minimization 
of $S_I$ leads to that this difference must have been adjusted 
- somehow or another - to be zero. 

This is the basis for the relation which is the main result of the 
present article, i.e. our prediction  comes from the imaginary action 
being minimized w.r.t. the collected neutron life time $T_{neutrons}$ and 
thus having 
zero derivative.     

Even if the above argument for the zero derivative of $S_I$ w.r.t. 
$T_{neutrons}$ and thus the change in $L_I$ under neutron decay 
being zero were not convincing from a completely theoretical point 
of view, we can at least note phenomenologically, that there is in 
nature today both neutrons and neutron-decay products, protons and electrons
(and (anti)neutrinoes) in comparable amounts. If indeed there were a 
difference in the imaginary part of the Lagrangian $L_I$ for the neutron
and its decay products then it should in our model go so 
that either all the neutrons disappeared or all the protons and electrons 
were made into neutrons. Neither of these two predicted possibilities 
seem to have been even approximately realized in nature. Thus our model 
will be in trouble unless the change in $L_I$ under neutron decay is indeed 
zero. With the enormous amounts of neutrons in the world one would 
get an enormous effect of selection of the history with an exponentially 
easily enormous factor.

\section{The interesting relation}

Now we shall evaluate the difference between the $L_I$-contribution 
of the two pairs of particles on the sides of the equation (\ref{wp})
because that is what we predict to be zero. 

Now we have that the main interaction with the Higgs field of some 
particle such as a quark or an electron simply comes via the Yukawa coupling 
which in turn is proportional to the mass of the particle in question.
Since we have hypotesised, that it is the square of the Higgs field, which 
gives us the $L_I$ dominant term, and we expect for small interactions 
that this square will vary proportional to the Yukawa coupling, we expect 
that the contribution from the passage, of a quark 
say, by multiplying the eigentime for the periode considered 
with the mass of the quark

Since a similar relation holds for the electron, say, we can thus see that 
for a non-relativistic electron the contribution to $L_I$ is simply 
proportional to the electron mass $m_e$. With same proportionality 
the contribution to the $L_I$ from a non-relativistic proton say 
will of course get three contributions one from each of the three 
quarks, but now these quarks are not non-relativistic but rather move 
most of the time pretty relativistically. Rather we have that the amount of 
eigentime spent per unit time in the rest frame of the proton is proportional 
$\gamma^{-1}$, where $\gamma$ is the relativistic $\gamma$-factor for the 
quark (in the proton system). Of course this eigentime correction 
factor $\gamma^{-1}$ fluctuates quantum mechanically and in reality has 
a distrubtuion rather than a special value. Nevertheless we expect that very 
crudely it behaves similarly to the inverse of the average of the 
$\gamma$-factor itself, and we have found it suitable to define a factor of 
order unity ``ln'' by the relation
\begin{equation}
<\gamma^{-1}> = ``ln''/ <\gamma>
\end{equation}       
where $<...>$ symbolizes the average over the quantum fluctuatuations of 
the quark in question in the nucleon, the proton say.  The reason we have 
chosen the symbol $``ln''$ for this correction factor of order unity is
that we expect that a contribution from the  part of the wave function or 
better distribution of the quark energy in which it is accidentally slow 
will cause a logarithmically divergent contribution to $<\gamma^{-1}>$
in the limit of the quark mass going to zero. There is therefore expected a 
term at least in $``ln''$ that goes like the logarithm of the constituent 
mass of the quark meaning really its energy divided by the quark mass.

To get an idea of what our $``ln''$ shall be we have to imagine some 
- rather smooth of course - distribution of $\gamma$ for the quark considered.
This distribution must have the average, so that 
\begin{equation}
<\gamma> = \frac{m_{const}}{m_q},\label{avg}
\end{equation}
 where $m_{const} = E_q$ is the energy on the average of the 
quark inside the proton, say. It is also essentially the constituent mass 
in the quark model. The current algebra quark mass - and the one involved 
in the Higgs-Yukawa coupling - of the quark is called $m_q$.

Now it is impossible to have $\gamma$ being less than unity, 
and we may use this fact together with a guess from smoothness of the
statistical distribution of $\gamma$ in the wave function of the quark 
inside the nucleon.

As a function of $\gamma$ of course $\gamma^{-1}$ is the inverse simply, and 
flat distribution would integrate up to give to $<\gamma^{-1}>$ essentially  
$\ln{(2<\gamma>)} /(2<\gamma>-1)$. Using this estimate in our definition of
$``ln''$ we would get the very crude estimate 
\begin{equation}
``ln'' \approx <\gamma>\ln{2<\gamma>}/(2<\gamma>-1) \approx 
\frac{ln(2<\gamma>)}{2 - <\gamma>^{-1}} \approx \frac{ln(2<\gamma>)}{2}
\label{est} 
\end{equation}
for big $<\gamma>$ as is indeed the case for quarks in nucleons. 

With for instance a down quark mass being $6 MeV$ while using for the 
consituent mass or better the energy for a quark in the nucleon as 
being one third of the full mass say take 330 MeV - or if one wants 
to give say half the energy to  non-valence, 165 MeV, 
we would then get 
\begin{equation}
<\gamma> \approx 330/6 = 55
\end{equation}  
and with this value 55  we get by (\ref{est}) 
\begin{equation}
``ln'' \approx \frac{\ln{(2 *55)}}{2 - 55^{-1}} =\frac{4.70}{1.98} = 2.37  
\end{equation} 
If we used instead the value in which half the energy is in the non-valence
quarks or gluons which are presumably the same for proton and neutron so that 
they do not contribute to the difference between them which is what we care 
for in the formula we are on the way to derive, we would get instead for 
the averga $\gamma$
\begin{equation}
<\gamma> \approx 165/6 = 27.5
\end{equation}  
and with this value 27.5  we get by (\ref{est}) 
\begin{equation}
``ln'' \approx \frac{\ln{(2 *27.5)}}{2 - 27.5^{-1}} =\frac{4.01}{1.96} = 
 2.05. 
\end{equation} 
(Had we used instead the small estimate of the down quark mass $m_d = 3.5 MeV$,
we would get the corresponding values for $``ln''$ as $2.634$ and $2.298$.)

Since the only difference between a proton and a neutron is the exchange 
of one of the up-quarks in the proton by a down quark  to make it a neutron,
the difference in the $S_I$ contribution is proportional to just the 
difference between the contributions for the quarks in quaestion. We thus 
obtain the equation needed to make the balance between the $L_I$ contributions 
on the two sides of equation (\ref{wp}) for which we argued a finetuning to 
occur:
\begin{equation}
m_e + m_u * <\gamma^{-1}>_u = m_d * <gamma^{-1}>_d. 
\end{equation}  
By insertion we get from this equation then:
\be
m_e + ``ln''_u \frac{m_u^2}{m_{const}} = ``ln''_d * \frac{m_d^2}{m_{const}},
\label{relation}
\ee
(where it is strictly speaking better to think of $m_{const}$ as the energy 
of a quark inside the nucleon)
or we can write it - assuming that the $``ln''$ not depending much on the 
quark 
\be
m_e = ``ln'' *\frac{m_d^2 -m_u^2}{m_{const}}.\label{rel}
\ee
Insertion of say 
\begin{eqnarray}
m_d & = & 3.5 MeV \hbox{ to } 6.0 MeV \\
m_u & = & 1.5 MeV \hbox{ to } 3.3  MeV \\
m_e & = & 0.51 MeV \\
m_{const}& = & 330 MeV \hbox{ or } 165 MeV\\
``ln''&  =& 2.63 \hbox{ or } 2.05
\end{eqnarray}
gives us 
the right hand side
\begin{equation}
r.h.s.(\ref{rel}) = 2.05 * \frac{36  MeV^2 - 11  MeV^2}{165 MeV} = 0.31 MeV
\end{equation}
One might wonder about possible corrections to this first estimate.
This agrees within 40 \% with the actual elctron mass, which is better than 
the accuracy of our estimate so far. 

It should be remarked though that, had we not made the assumption about the 
half of the energy being in non-valence partons, we would have got a worse 
agreement 
and predicted the electron mass about a factor 3 too low. Even that would 
though in the first run have been good enough.
 
Since indeed such a half of the energy being in the non-valence partons 
is phenomenlogically about right we should take the agreement of our formula
with experiment to be so good that it must be considered wellfunctioning. 

In principle - but perhaps not in pracsis - one might be able understanding 
and calculating with QCD or using phenomenological information to calculate 
much more precisely the average inverse $\gamma$ and the quark mass to be used
(e.g. the Yukawa couplings are running). By such a calculation we might hope 
in the future to check our prediction more precisely. The quark masses are 
not determined so accurately and it would be preferable if we could instead 
formulate our relation (\ref{relation}) as a relation involving the 
pion masses or the isospin breaking massdifferences directly. 

\section{How difficult to get the relation in other ways} 

It should be stressed that the here from the imaginary action model derived 
formula involves such quantities that it would be hard to see how it should 
come out of more conventional theory: In fact one could of course imagine that
we could have a theory connecting the quark and electron masses, because 
they would be involved with the physics behind the Yukawa couplings, but it 
seems difficult to see how the mass of the proton or the neutron could come 
in too. The mass of the nucleons and thus the energy of the quarks in 
the nucleons are namely given mainly by the QCD-$\Lambda$. This QCD -lambda 
must be extremely sensitively depending on the physics at the presumably very 
high energy scale at which the Yukawa couplings presumably get their 
values determined. Thus it would be rather accidental, if our formula should 
be obeyed for some other physical reason. So it would either be our 
derivation, or it would be just accidental.   

\section{The Cosmic Ray problem}
It has been claimed that our model prediction of production of
Higgs particles causing ``bad luck'', meaning that such production should 
be prevented is already falsified by the fact that there are cosmic rays 
hitting the earth with such energy that certainly Higgses should be produced 
according to the Standard Model. It should, however, be understood that our 
model does 
{\em not} simply mean that such production will not occur at all, but 
rather that such production is potentially allowed to some extend
provided it pays with respect to minimizing the ``imaginary part 
of the action'' $S_I(history)$. That is to say, that, if there were e.g. a
mechanism for production of cosmic radiation,of  which it would be almost 
impossible to get rid, by almost any ``bad luck'' however cleverly 
arranged, then such production would have to be there, basically because 
it cannot be prevented, unless the $S_I$-contribution from say big bang 
eras would have to be increased dramatically. The true prediction of our 
model will thus only be that Higgs production is brought to be so low as it can
be {\em within the possibilities reachable without increasing $S_I$ in some 
other era such as having Higgs production in an other era}.

In spite of our prediction being in this sense less strong than a total 
prevention of Higgs production we might still expect, that there would be some 
observable reduction of cosmic rays in the energy range that could lead to 
Higgs production. In the cosmic rays the amount of protons dominates over the 
amount of antiprotons and thus the Higgs production will be dominated 
by gluon collisions so that the Higgs production effectively only begins 
when the gluonic parton distribution function upper (approximate) edge 
reaches so high in energy that the Higgs with whatever mass it now may 
have get producable. 

If we e.g. take the edge of the gluon distribution to lie at 0.1 and think 
of Higgs with mass 120 GeV then the effective threshold for Higgs production
becomes at $\sqrt{s} = 120GeV/2/0.1 = 600 GeV$, which in turn comes to mean 
that the beam energy $E_{threshold \ Higgs}$ needed for production
of Higgses is given by
\begin{equation}
E_{threshold \ Higgs} = \frac{s_{threshold \ Higgs} - 2m_p^2}{2m_p} = 
\frac{(600GeV)^2 - 2 GeV^2 }{2 GeV} = 1.8 *10^5 GeV = 1.8 *10^{14} GeV.
\end{equation} 

Interestingly enough it now happens, that order-of-magnitudewise this 
threshold for Higgs production is very close to the already wellknown ``knee''
\cite{knee} at which the curve of the intensity of cosmic radiation 
as a function of the 
energy bends downward, so that indeed there is approximately a stop for 
the cosmic ray very roughly just at this Higgs producing threshold.
The explanation\cite{explanation} for this ``knee''  \cite{knee,knee2} is 
presumably that the supernovae
in the  galaxy can only produce protonic cosmic ray up to this ``knee''-energy
of the order of $10^{15}$ GeV; then it may still be possible to have from 
this supernovae source some higher-Z nuclei with energy above this ``knee''.
Indeed phenomenological evidence is that above the knee the cosmic ray 
particles are dominantly Fe-nuclei\cite{knee2}.
  According to e.g. the plot in the article by Thomas K. Gaisser
\cite{Gaisser}  arXiv:astro-ph/0608553v1 there is a  ``knee'' at the energy 
$~ 2 *10^6 GeV = 2*10^{15} eV$. This is very much where we 
like to have it in order to  just barely avoid the Higgs production.

One would almost say that the appearance of the ``knee'' just at this
place  - on an order of magnitudewise even very long curve of various energy 
scales having been investigated for cosmic rays - is almost remarkably 
good. So the knee should be considered a victory of our model!

In the philosophy of our model we should consider this closeness of the 
``knee'' with the Higgs production threshold as not accidental, but rather
e.g. the parameters or coupling constants or some details of the history 
have been adjusted, so that the highest cosmic ray energies achievable 
by supernovae comes to be very close to the Higgs threshold. A priori it is 
only the initial conditions we have suggested to be fixed by the minimization 
of the ``imaginary part of the action'' $S_I(histoty)$, and that could imply
that for instance the Hubble expansion rate could be what gets adjusted,
but it seems to be a very atracktive idea to allow the adjustment towards 
minimizing $S_I$ not only to concern initial conditions, but also the coupling 
constans such as we have just seen above. 

\subsection{More accurate estimate of the effective Higgs production threshold}
Since the ``knee'' happens to be so close to the effective threshold for Higgs 
production, it becomes interesting to define and estimate this Higgs-production
threshold a bit more accurately. There is not truly any threshold for 
Higgs production in the range of energies, where there is any significant 
chanse for producing Higgses at all. For instance for the Higgs mass expected 
in the type of model connected with the present article 120 GeV the formal 
Higgs threshold in terms of $\sqrt{s}$ would only be a couple GeV more than 
the mass 120 GeV of the Higgs, since it would only be needed to have in 
addition to the Higgs two protons for baryon number and chanrge conservation. 
At 
this energy there is, however, no Higgs production at all in practice.Indeed 
the Higgs production cross section has a very tiny tail at low energy 
between the formal threshold and a much higher square root $s$ closer to 
the Tevatron or LHC energies. Let us therefore define for each mass 
possibility for the Higgs $m_H$ an ``effective Higgs production threshold'' 
by a {\em linear} extrapolation from above in energy (above  the insignificant 
tail) of the Higgs production cross section. In fact such a linear 
extrapolation of the Higgs cross section in the energy range where there is 
a significant production will go to zero at some finite $\sqrt{s}$, and that 
value where this linear extrapolation goes to zero we shall call the  
``effective threshold''. Using the theoretical prediction  curves from 
the article by Tully \cite{Tully} we obtain for instance using 
a Higgs mass of $m_H = 150 GeV$ that the cross section at $\sqrt{s} = 30 \ TeV$
is $10^{-1} \ nb$,  at $\sqrt{s}= 7 \ TeV$, $\sigma = 1.5 *10^{-2} \ nb$ 
while
at $\sqrt{s} = 14 \ TeV$ it is $4 * 10^{-2} \ nb$. From this we get by 
linear extrapolation a zero of the cross section for Higgs mass $150 \ GeV$ 
about $3.4 \ GeV$. Similarly we could  for a Higgs mass of $500 \ GeV$
find cross section $2.5 * 10^{-2} \ nb$ at $30 \ TeV$, and $10^{-3} \ nb $
at $7 \ TeV$, while it is $5 * 10^{-3} \ nb$ at $14 \ TeV$. This gives 
us - using to fit linearly only the 14 and the 30 TeV points  -
an extrapolated zero at $\sqrt{s} = 10 TeV$.

Theoretically we expect the effective Higgs threshold to scale in $\sqrt{s}$ 
 proportinally to the Higgs mass - assuming the parton distrubtion 
functions to be rather constant - and so it is comforting that the about 
a factor three in mass between 150 GeV and 500 GeV matches with a ratio 
close to three for our crudely estimated thresholds 3.4 TeV and 10 TeV
for the two masses respectively. For our favourite low mass of 120 GeV - also
the one favored by the indirect measurements and the masses more and more 
getting excluded higher up - we extrapolate to an effective 
threshold  2.7 TeV. Thus we arrive at the an estimate 
for effective Higgs production threshold for Higgses of a small mass 
say arround 120 GeV as is the main left over range of Higgs mass 
is ca 2.7 TeV in $\sqrt{s}$. This would correspond to a fixed target beam 
energy $ (2.7 *10^3)^2 /2  \   GeV = 3.6 *10^6 \  GeV = 3.6 *10^{15}\ 
eV$. This is very close indeed to the value $2 * 10 ^{15} \ eV $ which
we extracted from the Gaisser curve above. So indeed the agreement 
with the by linear extrapolation defined threshold for a ``light Higgs'' 
is very close to the ``knee'' !

\section{Estimation of the crucial number of Higgses to produce backward 
casation}

In addition to the above mentioned problem of the cosmic ray Higgs production 
our complex action model also has the problem that the Tevatron at FNAL in
Chicago (Batavia) presumably already has produced about say 10000 Higgses.
Truly we do not know, if it has, because no Higgs bosons have been 
convincingly observed so far (i.e. in early 2010) and the mass of the 
Higgs is also known only through very uncertain indirect messurements.
But if as is actually supported through a model supported by the picture
connected with the complex action model of the present article \cite{degenerate}
\cite{Higgsmass} the Higgs mass is equal to the lower bound for it in the
Standard Model, then there would have been already produced according to 
the Standard Model several thousands Higgs bosons in the Tevatron. 
With such a low mass of the order of $120 GeV/c^2$, however,. even several 
thousands of Higgs produced would not have been seen yet. 

If we shall uphold the model that Higgs particle production cause bad luck 
for the production machine we have therefore to withdraw to the position, that 
it is only a sufficiently big number of Higgs bosons being produced, that will
cause sufficient effect to truly cause some visible change in the chance 
of the number being produced indeed.

\subsection{How does the probability for bad luck depend on the number of
Higgs bosons produced}

A very naive and simple thought in our model gives immediately that as the 
Higgs boson 
living a time in its rest frame just on the average equal to the 
Higgs-life-time  the suppression of the amplitude 
(\ref{amplitude}) occurs with a factor being
\be
(exp(-L_I(Higgs)\tau_{Higgs}))^{\#Higgses} = K^{\#Higgses}. 
\ee       
Here $L_I(Higgs)$ is the contribution in the rest system of the Higgs from
the Higgs particle, and $\tau_{Higgs}$ is the average life time form for the
Higgs particle, while $\#Higgses$ denotes the number of Higgs bosons 
produced. Then the probability which goes as the numerical square this 
amplitude will also go with the 
number of Higgses $\#Higgses$ in the exponent, $(K^2)^{\#Higgses}$.

This simple way of looking at it ignores the effect of the competion 
between different eras in governing the initial conditions. By the competion 
with the other eras 
 the dependence of the intial conditions or equivalently the realized 
solution on what goes on in a given era (our era say)is borought  appreciably 
down. It is,
however, expected that inclusion of this era-competition-effect will
change the constant $K$ to be much closer to unity. If we therefore 
just define a phenomelogical constant 
$a = -\ln{(K^2)}|_{after \;\; era-competion-correction}$  
we can crudely estimate the probability change to the usually expected 
probability distribtuion to say a card pull due to it being made responsible
for the switch on or not of a Higgs-producing macine, such as LHC say.

Let us in fact imagine that we decided to pull a card from a usual
card-deck and to let LHC be stopped, if we pull a black card while it
gets allowed to run fully, if the card pulled is red. Then if our model 
of imaginary part in the action were not true, there would of course 
be 1/2 probability for red and 1/2 for black. If now our model were right,
however, the non-normalized probability for the red card would be suppressed 
by a factor $exp(-a\#Higgses)$ relative to what it were without 
the effect of our imaginary part of the action. After normalization we 
would then get 
\begin{eqnarray}
probability(red)& = & \frac{1/2}{1/2 + 1/2 *exp(-a \#Higgses )}= 
\frac{1}{1+exp(a\#Higgses)} \\
probability(black) & = & \frac{1/2 * exp(-a \#Higgses)}
{1/2 + 1/2 *exp(-a\#Higgses)} =  \frac{1}{exp(a\#Higgses)+1}\\
\end{eqnarray}    
  
It is from these expressions clearly seen, that one {\em only gets 
a significant effect of the imaginary part of the action provided
the product $a\#Higgses$ is of order unity or bigger}. Thus the 
``phenomenological parameter'' $1/a$ becomes approximately the number 
of Higgses needed to produce in order to give any significant 
effect via our model. It is easy to see, that with usual or start 
probabilites $P_s(red)$ and $P_s(black)$ not equal the effect gets less easy 
to observe. In fact    
  \begin{eqnarray}
probability(red)& = & 
\frac{P_s(red)}{P_s(red) + P_s(black) *exp(-a \#Higgses} \\
probability(black) & = & \frac{P_s(black) * exp(-a \#Higgses)}
{P_(red) +P_s(black) *exp(-a\#Higgses)} \\
\end{eqnarray}
is calculated by correcting by the suppression factor and normalizing 
again to the total probability being unity.

\subsection{Estimation of the ``critical'' amount of Higgses $1/a$ needed 
to cause any backward causation}
From the well-running of the Tevatron so far (2010) we can conclude that 
the parameter $1/a$ should not be terrible much smaller than the about 
10000 which we may take as the order of magnitude for the number of Higgses 
having been  produced by this Tevatron already. It can though be somewhat 
smaller 
since it is possibly difficult by adjusting the initial conditions to 
get such a machine prevented from getting built or to come to run.

Most important for saying something about our parametrization $1/a$ of our 
effect is to see what we deduce about it by believing that the failure 
of SSC were indeed due to our effect. This must mean that indeed $a\#Higgses$
were at least of order unity, but presumably preferably bigger than unity,
where $\#Higgses$ is taken to be {\em the number of Higgses that would have 
been 
produced in this SSC-accelerator.} According to one plan there is projected 
a development through 12 years of running (it is of course very difficult 
to know even for how long a successful SSC-accelerator would have been 
allowed to work in the hypotetical case, that it were not killed before 
working at all, but let us here for the etimation take these 12 years
as a good estimate.). It should then have had luminocities starting at the 
first 0.5 year as $10^{31} cm^{-2}s^{-1}$, becoming at year 1 
$10^{32}cm^{-2}s^{-1}$, at year 2 a luminocity $10^{33} cm^{-2}s^{-1}$,
at year 5 luminocity $2.8*10^{33} cm^{-2}s^{-1}$, reaching at year 10
the luminocity $8 * 10^{33} cm^{-2}s^{-1}$, ending with in the 12th year
$10^{34} cm^{-2}s^{-1}$. Say that it from this would work in about 
the last 5 years with a little less than the lumninocity 
$10^{34} cm^{-2}s^{-1}$; it would say with a high proportion of time 
being working indeed have produced an integrated luminocity of the order 
a bit less than $5 * 3* 10^7 s  * 10^{34} cm^{-2}s^{-1}$ =
$1.5 *10^{42} cm^{-2}$. Let us say $10^{42} cm^{-2} = 10^{18} barn^{-1} =
10^3 fb^{-1}$. 

Correspondingly one has for LHC thought of an integraget luminocity up to 
2025 of $5028 fb^{-1}$, only deviating by a factor 5 from the expectation 
for SSC, LHC having w.r.t. integrated luminocity only a bit (a factor 5)
bigger expectation than SSC. 

At LHC at full energy $7TeV + 7TeV = 14 TeV$ and Higgs mass 120GeV a cross 
section of the order of $30 pb$ is expected for Higgs production.
Thus at LHC if mainly running  full energy  we get {\em produced}
- but certainly not all observed - $5* 10^3 fb^{-1}* 30 pb = 1.5 *10^7$ 
Higgs particles. At SSC the production would be somewhat bigger because of a 
higher cross section at the Hihger top energy $20 TeV + 20TeV = 40 TeV$.
Even a factor 9 bigger crosssection would only lead to similar order of 
magnitude of the total number of Higgses produced as in LHC, because of the 
higher luminocity of LHC, say a factor two more Higgses in all than at LHC.

If indeed LHC should ever come to produce more Higgses than SSC would have done
- which though does not sound so easy according to the just given estimates -
then the best evidence for our model, the failure of SSC would be lost and 
our model would, if not formally, then in pracsis be falsified. 

\subsection{An estimate of the parameter $1/a$ for the crucial number of 
Higgses}
However, we think we may get a true estimate of the order of magnitude 
for the significant number of Higgses $1/a$ by making the assumption that 
it were not just accidental that at the moment, March 2010, LHC is running at 
$3.5 TeV + 3.5 TeV = 7TeV$ rather than the schedule for full 
energy per particle $7TeV + 7TeV = 14 TeV$, because of the physicists 
having been scared by troubles caused by the ``God'' in our model.
If we in fact assume that there were organized - by our model initial 
conditions - some troubles connected with the incident more than one year ago, 
when there were the explosion in the tunnel causing, that for safety 
the LHC in this moment has to run at the half originally planned energy 
rather than at this full energy $14TeV $ in center of mass, then it would mean
that the ``God'' so to speak would {\em care for the Higgses potentially 
being produced in the supposedly 18 month periode concerned, if it 
had ran with the high energy per particle}. Actually we should more
precisely say that this ``God'' would have to care even for the difference in 
numbers of Higgses produced at $7 Tev + 7 TeV= 14 TeV$ and  the 
number produced at only $3.5 TeV + 3.5 TeV = 7 TeV$. But hat does not 
matter so 
much order of magnitudewise, because the number of Higgses at 
$3.5 TeV + 3.5 TeV = 7 TeV$ will be at least a factor 2 smaller than at 
the full energy, and thus the difference in number of Higgses will be 
order of magnitude of what is produced at the full energy of 
$7 TeV + 7TeV = 14 TeV$. That is to say order of magnitudewise 
we may estimate also the difference to represent about one or one and 
a half year of beginning Higgs production at full energy having the 
$30 pb$ as typical crosssection.    
         
   According to some old expectations one should have in LHC $6 fb^{-1}$
in 2009. Now let us interprete it to mean first running year and twice as 
much second year, which would integrate up to integrated luminocity 
after two first years of $18 fb^{-1}$. It is presumably crudely o.k. then 
to think of $10fb^{-1}$ as the integrated luminocity at the end of the 
presently going 18 month periode of running in 2010 to 2011. 

The estimated number of would have been with full energy production 
in these first 18 month would thus be $30 pb * 10 fb^{-1} = 3 * 10^5 $
Higgses. We therfore would say that under this assumtion of the 
scaring of the physicists to only work with half energy at the moment 
means that crucial number of Higgses $1/a$ should be of the order
of this number $3 *10^5$. 

If this 
\be
1/a \approx 3*10^5 
\ee 
is indeed true then the SSC with its say $3*10^6$ Higgses, it is 10 
times as many, would almost certainly have to get stopped somehow 
if it were at all possible. It would namely have a suppression 
factor of the order
\be
``SSCsuppresion'' \approx exp(-a * 3* 10^6) = exp(-10). 
\ee 

It should be stressed that this last case of ``God'' scaring the energy 
per particle down to the half value gives roughly the order of magnitude 
for the $1/a$ parameter, becuase we truly have two number of Higgses being 
tested off - the high number not being allowed, and the one of the half 
energy being alllowed, at least if LHC truly comes to run for of the order 
of the 18 months planned - and so it gives both upper and lower limit.

We thus really have at least this very weak argument in favour of $1/a$ 
truly being of the order of 300000 Higgses.

So if some day LHC reaches to have produced appreciably more than these 300000
Higgses, then, although we could still screw up a bit our $1/a$ to accomodate 
such a for our model bad happening, we would then loose the evidence we could 
now say we see for our model by the scaring down to half  energy.
So if LHC comes to produce more than these 300000 Higgses then we would loose
 this case of evidense and it would be so bad for our model, that we should 
essentially consider it a falsification of our model.

This falsification the just suggested way would occur about the 18 months 
after the restart after the updating to $\sqrt{s} = 14 TeV$ after the shut 
down after the present 
18 months of running at the ``low''  $\sqrt{s}=7TeV$. AT that time we can 
write the paper about our model having been 
falcified. (I have been suggested this idea of having a falcification
 by Boerge Svane Nielsen)
  
\section{Conclusion}
We have reviewed a model by Ninomiya and myself based on the assumption of 
the fundamental theory being given by the Feynmann-Wentzel-Dirac path way 
integral in an interpretation involving the time at all times including both 
past and future; {\em But most importantly we take the action to be complex!}
 In a suggestive approximation the model of ours, which 
turns out to be a model also for the initial conditions, suggests that the 
(true) solution of the classical equations of motion, through which we live,  
should be selected by the requirement of minimizing the imaginary part
of the action $S_I[history]$, so that the latter  is the smallest achievable 
for 
just this true history of the universe. This approximation may be described 
by an analogy to a skier with frictionless skies which  after severe 
computer simulation calculations is being started with just the right speed 
and 
direction etc. so as to come through the according to the calculation 
 absloutely most beautifull (when integrated up over time) tour.
The integrated amount of beauty is in the analogy what corresponds to 
the imaginary part of the action $S_I(history)$. When the tour is constructed 
to completely optimize/maximize the integrated (amount of)beauty it is 
the tour
analogous to the realized history of the universe (with its analogously 
minimized $S_I(history)$). We should  then find, if our model is true, 
that the likely features of such an optimized tour on ski can be very similar 
in an abstract way to what, we see our history of the Universe to be 
phenomenolgically: If there were some place in the landscape that were so 
beautifull that any optimization hardly could avoid making the tour pass that 
spot, then the optimized tour would indeed pass that spot; but of course 
most likely the very most beautifull place would lie on a very steep 
hill side and it would not be possible to remain there for long with 
frictionless skies. This most beautifull place should be analogous 
to some time arround the ``big bang time''. If this biggest beauty 
region had so great beauty that it would dominate almost all the rest,
we could get in the analogy that the big bang time would have a rather 
definite state, but that what goes on in other eras will then essentially 
be a consequence of what were to be organized in this crucial era arround 
big bang, say to have on the to be realized history some special 
inflaton-field over an optimized long time. 
This picture favours a sort of bouncing-universe picture with the most 
important  era ``Big Bang time'' - which does presumably not have a genuine 
Big Bang but only that the Universe were very small, but not necessarily zero, 
size - determines to a very large extend what goes on both before and after
this era. We should then have in mind that, had we lived in the time 
before the ``Big Bang'' era, we would presumably have swichted our notation 
under a timereversal symmetry and changed the notation from the story, that 
we live in a world preorganized to reach a special Big Bang situation 
with an extremely negative $L_I$ and that we see order comming up 
(meaning entropy falling).It would have looked that things were getting 
prearranged  so as to make this great event of the inflation 
like periode be possible with the extremely negative $L_I$. But we would 
easily have timereversed the notation and in stead told the usual 
story that the inflation with the specially low (presumably very negative) 
$L_i$ would be called to be in the past rather than in the future.
In this way we could always interprete what happens by saying that on our 
side of the big bang era entropy grows and universe expands.

It has earlier been argued that the very likely the most important term
in the imaginary part of the lagrangian  
- by a factor $10^{34}$ we suggest - is the one comming from the 
imaginary part of the coefficient in the Higgs mass term, i.e. the term
proportional to the Higgs field square. It has been remarked that once 
we identify this term as dominant  there is no more so many parmaters 
effectively in the imaginary part as in the real part, because now we can 
ignore all the many small terms in the imaginary part $S_I$ and only care 
the one term proportional to  the Higs field square. This fact gives us better 
chanse for getting predictions of a less general character.We namely know 
the $S_I$ effectively under Standard Model conditions up to an over all 
coefficient.

For instance we can under the assumption of a limit of many parameters 
argue that there should be a minimum of $ S_I$ as a function of how 
the baryons are distributed (averaged over time) between protons and neutrons. 
The argument that it should 
indeed be so may not be quite watertight, but it could very likely happen 
that the direct effect of the number of baryons being there as protons versus 
those that are neutrons could become a significant term in the imaginary 
part of the action $S_I$. This would have caused there to be either almost no 
protons or almost no neutrons. To avoid such an incorrect scenario  we 
reached the necessity of having a 
certain 
relation between the light quark masses and the electron mass. This relation 
 (\ref{rel}) may also be written in the form
\begin{equation}
\sqrt{m_d^2 -m_u^2}= \sqrt{E_q* m_e}/\sqrt{``ln''} .\label{relcon}
\end{equation}
where we have used the notation $E_q \approx m_{const}$ to stress that it is 
really the energy of the quark $E_q$ in the frame of the nucleon that goes 
into our formula to give the $\gamma$ for the quark which is the imortant 
thing.
We used the energy of the valence quark being 1/6 of proton mass 1GeV and 
took 165 MeV, which then gives using for the order unity quantity 
$``ln'' \approx 2.05$ that we predict 
\be
\sqrt{m_d^2 -m_u^2} \hbox{`predicted to''} 9.17/1.43 MeV  = 6.41 MeV 
\ee
to be compared to 
 \be
\sqrt{m_d^2 -m_u^2} =\sqrt{3.5^2 -1.5^2} MeV \hbox{ to } 
\sqrt{6.0^2 - 3.3^2}MeV =  3.16 MeV \hbox{ to } 5.01 MeV.
\ee

We stressed that taking this relation as a success can give a rather strong 
support for our model with imaginary part also for the aaction..
It would namely 
be very difficult for any competing theory to reproduce this 
relation, because it involves the QCD-scale,and it is really hard to see 
how that
could be connected to the quark and electron masses in such a way
as our relation (\ref{rel},\ref{relcon}) states. So our relation should be 
considered 
a support 
for our model of imaginary action with its prearrangements. 

We also discussed the important problem with the cosmic rays, that should 
not hit the earth or other astronomical objects, if really Higgs production 
should be prevented. Of course it might be so difficult to switch off 
fully the high energy cosmic ray, so that a total cut off of the high energy
spectrum  would not pay in the attempt to minimize $S_I$.
 Very interestingly in this connection there is,
however, as  it has long been known  at about the energy of the cosmic 
ray particle $10^{15} eV$, {\em which happens to be very close to the 
effective threshold for producing many Higgs particles }, a rather
sharp fall off, more rapidly than at lower energies. This is what is known as 
the ``knee''
in the cosmic ray spectrum. As this spectrum has been studied well over several
orders of magnitude, even the only order of magnitude coincidence of the 
energy scale at which Higgses begin to be produced copiously and the scale 
of the ``knee'' becomes somewhat remarkable! Did our imaginary action 
model indeed arrange that the main source of cosmic ray from supernovae 
in the galaxy just stops, where the Higgses begin to be produced copiously ?
The effect of this ``knee'' in suppressing the production of many Higgs bosons 
in hits on astronomical objects is further enhanced by the result \cite{knee2}.

\subsection{Further study}
Some of the fine tunings being possibly explained by some antropic principle 
derivation, might likely be instead explained by the $S_I$-minmization.
For instance the existence of stable Helium-2\cite{diproton}\cite{McKee} 
nucleus could 
potentially 
increase the rate of stardevelopment appreciably by making the weak 
interaction proton-proton 
process starting the nuclear formation process in the stars be replaced by 
the electromagnetic formation of Helium-2. If there were such an effect 
the fact that the Helium-2 nucleus is just very barely unstable and thus 
useless in the star development could be considered a result of our model.
In fact our minimization of $S_I$ would be seeking to delay the formation 
of black holes with lot of high energy physics going on such as e.g. Higgs
production. Now, however, Bradford\cite{Bradford} has pointed out that 
in contrafactual 
world with a stable Helium-2  stars with similar 
life times as they have the real world are not excluded. So the argument 
is not neccesarily so simple; but in our model whenever something dramatically 
would happen by varying a coupling there is a high chance that it would cause 
alssos dramatic effects on $S_I$ and at the end drive the history of the 
universe and probably even the couplings to adjust into neighborhood of the 
dramatic shift.   

\section{Acknowledgement}
We would like to thank John Ellis for stressing the problem with the cosmic ray
provoking the looking for the ``knee'' about which I were also told in the 
airport back from Spaatind Conference by one of the participants. 
Also I thank Keiichi Nagao with whom I have a running project on the 
harmonic oscillator etc. in the complex action model. Most of the 
work were of course a review of work with Masao Ninomiya as described,
a collaboration for which I also thank. 
I am thankful to Angeliki Koutsoukou-Argyraki both for discussion and for 
reading the manuscript and 
and finding the reference\cite{knee2}. I thnk Svend Erik Rugh and 
Thomas D{\o}ssing for informing me about the Helium-2 stability 
discussion. 
Further  
I thank 
FNU for supporting the trip to Spaatind, Norway,for the conference where I 
delivered a talk mainly on the review part of this article.


\end{document}